\documentclass[prd,aps,superscriptaddress,floatfix,nofootinbib,eqsecnum,onecolumn]{revtex4-1}
\pdfoutput=1

\usepackage[dvipsnames]{xcolor}
\usepackage{amsmath}
\usepackage{amssymb}
\usepackage{physics}
\usepackage{bbold}
\usepackage{graphicx}
\usepackage{verbatim}
\usepackage{hyperref}
\usepackage{empheq}
\usepackage{float} 
\usepackage[normalem]{ulem}
\usepackage{subcaption}

\usepackage{placeins}

\usepackage{indentfirst}

\begin{document}

\title{Back-Reaction of Super-Hubble Fluctuations,  Late Time Tracking and Recent Observational Results}
\author{Marco Antonio Cardoso Alvarez}
\email{marcocardoso@id.uff.br}
\affiliation{Instituto de Fisica, Universidade Federal Fluminense, 24210-346 Niteroi, RJ, Brazil}

\author{
Leila Graef}
\email{leilagraef@id.uff.br}
\affiliation{Instituto de Fisica, Universidade Federal Fluminense, 24210-346 Niteroi, RJ, Brazil}

\author{Robert Brandenberger}\email{rhb@physics.mcgill.ca}
\affiliation{Physics Department, McGill University, Montreal, QC H3A 2T8, Canada}


\begin{abstract}
Previous studies suggested that the back-reaction of super-Hubble cosmological fluctuations could lead to a  dynamical relaxation of the cosmological constant. Moreover, this mechanism appears to be self-regulatory, potentially leading to an oscillatory behavior in the effective  dark energy. Such an effect would occur in any cosmological model with super-Hubble matter fluctuations, including the standard $\Lambda$CDM model.

The recent DESI data, which indicate that dark energy may be dynamical,  has renewed interest in exploring scenarios leading to such oscillatory behavior. In this work we propose a parametrization to account for the impact of super-Hubble fluctuations on the background energy density of the Universe.  We model the total effective cosmological constant as the sum of a constant piece and an oscillating contribution. We perform a preliminary comparison of the background dynamics of this model with the recent radial BAO data from DESI. We also discuss the status of the $H_0$ tension problem in this model. 

\end{abstract}

\maketitle

\section{Introduction}%
\label{sec1}



In the standard framework of cosmology, the $\Lambda$CDM model, the dark energy (DE) component, which comprises the majority of today's cosmological energy density,  is parametrized by a positive cosmological constant term, $\Lambda$, in the Einstein equations. Despite its apparent simplicity, this interpretation faces conceptual problems. In addition to some well known problems,  such as the cosmological constant problem,  the coincidence problem and the Hubble tension problem, several works in the literature have been pointing to an instability of a de Sitter phase.   In particular, there are indications that a de Sitter space could be unstable due to infrared (IR) effects \cite{Polyakov:2012uc, Polyakov:2007mm, Mazur:1986et, Mottola:1985qt, Mottola:1984ar, Tsamis:1994ca} \footnote{In the context of String Theory there have also been indications that it is not possible to obtain a stable de Sitter phase \cite{Obied:2018sgi, Dvali:2014gua,Danielsson:2018ztv,  Dasgupta:2018rtp},  and the recently suggested ``Trans-Planckian Censorship Criterion'' \cite{Bedroya:2019snp, Bedroya:2019tba, Brandenberger:2021pzy} implies that a positive cosmological constant is not consistent from an effective field theory point of view.}.   Already in the work of Ref.  \cite{Tsamis:1992sx, Tsamis:1996qq} it has been suggested  that the reason that our universe is not inflating today is that IR processes in pure quantum gravity (no matter) tend to screen the bare $\Lambda$. However, the authors of \cite{Tsamis:1996qq} found that  these effects become important only after perturbation theory has broken down. In 1997, the work of Ref.\cite{Abramo:1997hu} studied the back-reaction of super-Hubble modes of cosmological perturbations induced by matter, finding that the effective energy density associated with these modes would counteract any pre-existing $\Lambda$ (see also \cite{Finelli:2001bn, Finelli:2003bp, Marozzi:2006ky}). Later, in 1999, in the work of \cite{Brandenberger:1999su} it was speculated that this dynamical relaxation mechanism for $\Lambda$ could be self-regulating \footnote{Note also the related ``Everpresent Lambda'' scenario, which was proposed by  drawing on ideas from causal set theory and unimodular gravity, and which leads to an effective  fluctuating cosmological “constant”, see Refs.\cite{Ahmed:2002mj, Das:2023hbw, Das:2023rvg}, for instance.}. A crucial question that remained at that time was whether this back-reaction effect could be locally  measured \cite{Unruh:1998ic, Geshnizjani:2002wp, Abramo:2001dc}. Recent works have shown that super-Hubble fluctuation modes can in fact modify the parameters of a local Friedmann cosmology, when measured by a clockfield \cite{Geshnizjani:2003cn, Marozzi:2012tp, Finelli:2011cw, Gasperini:2009wp, Gasperini:2009mu, Marozzi:2010qz, Marozzi:2011zb, Abramo:2001dd, Losic:2005vg, Losic:2006ht, Brandenberger:2004ix, Afshordi:2000nr, Comeau:2023gxk, Comeau:2023euf} \footnote{In the context of the back-reaction of sub-Hubble cosmological fluctuations, a recent analysis  can be found, for instance, in Ref.\cite{Giani:2024nnv}}\footnote{Another interesting approaches to account for the back-reaction effect can be seen in the works of Refs.\cite{Kolb:2009rp,Kolb:2011zz,Marra:2011ct}, for instance.}. In the work of \cite{Brandenberger:2018fdd} the analysis was extended beyond perturbation theory, showing that the locally measured Hubble expansion rate indeed obtains a negative contribution from the back-reaction of super-Hubble fluctuations, whose amplitude grows in time. This supports the suggestion that (quasi) de Sitter space-times are unstable, and that this back-reaction mechanism  might lead to a dynamical relaxation of the cosmological constant. As first speculated in Ref.\cite{Brandenberger:1999su}, this is expected to lead to an effective dark energy density which oscillates in time around the dominant matter density.

The quest for  understanding the nature and dynamics of the  accelerated expansion of the Universe became even more interesting recently with the improvement in the measurements of the expansion history, which are increasingly allowing us to constrain the expansion history of the Universe. Recently, the Dark Energy Spectroscopic Instrument (DESI) Year 1 (DR1)\cite{DESI:2024aqx, DESI:2024kob, DESI:2024mwx} and DESI DR2 \cite{DESI:2025zgx, DESI:2025fii} results were released,  officially marking the start of the Stage IV DE era. The cosmological results stem from measurements of baryon acoustic oscillations (BAO) in galaxy, quasar and Lyman-$\alpha$ forest tracers. The DESI results, when combined with a number of external probes, provide intriguing hints for a dynamical, time-evolving DE component \cite{DESI:2024mwx, DESI:2024aqx, DESI:2024kob}. These hints were present already in the DR1 release \footnote{See, however, \cite{Payeur:2024dnq} for caveats.}. Later, the enhanced statistical power and extended redshift coverage of DESI DR2 sustained the preference for a dynamical dark energy component, yielding a level of statistical significance comparable to, or greater than, that reported with DESI DR1 \cite{DESI:2025wyn,DESI:2025fii, DESI:2025zgx}.  These hints may have enormous implications for the understanding of the nature of DE, and align with  expectations from certain fundamental theory  frameworks \cite{Brandenberger:2025hof}. 

In light of these results, recent works, among them \cite{Jiang:2024xnu} and \cite{Escamilla:2024fzq}, have pointed to signals, from late-time cosmological data, of oscillatory/non-monotonic features in the shape of the expansion rate at low redshifts \footnote{See,  however, \cite{Akarsu:2022lhx} for a warning that oscillatory features in $H(z)$ can be induced when fitting for this function given incomplete data. Also, the DESI full shape modelling does not show a signal \cite{Colgain:2024mtg}.}. The possibility of an oscillating dark energy has already been addressed  in earlier works, as for instance in Refs. \cite{Zhao:2017cud,Tamayo:2019gqj,Rubano:2003er,Linder:2005dw,Feng:2004ff,Nojiri:2006ww,Kurek:2007bu,Jain:2007fa,Saez-Gomez:2008mkj,Kurek:2008qt,Pace:2011kb,Pan:2017zoh,Panotopoulos:2018sso,Rezaei:2019roe,Yao:2022jrw,Rezaei:2024vtg, Tian:2019enx, DESI:2024kob, Adil:2023exv, Mbonye:2022cnf, Mbonye:2024mss}. In Ref. \cite{Escamilla:2024fzq}, in light of the recent cosmological data, including DR1, twelve different parametrizations for an oscillatory dark energy equation of state were analyzed,  among them the ones proposed in Refs. \cite{Linder:2005dw,Feng:2004ff,Pan:2017zoh,Zhao:2005vj}. This analysis demonstrated that all the oscillating DE parametrizations considered provide a better fit to the DR1  when compared to the cosmological constant, although in the Bayesian evidence analysis they are penalized  due to the additional free parameter. In a similar context, in the work of Ref. \cite{Jiang:2024xnu}, the late-time  expansion history was reconstructed in a way,  according to the authors,  to be ``as model-independent and non-parametric as possible" (see also \cite{Heisenberg:2020ywd,Haude:2019qms}). In their first reconstruction, using DR1 + PantheonPlus data,  two features in the unnormalized expansion rate were identified, a bump at redshift $z \sim 0.5$, and a depression at $z \sim 0.9$  (both features  were absent when replacing the DESI  dataset with the older SDSS one). 
On the other hand, the depression feature was noticeable in both the DR1+PantheonPlus and DR1+Union3 reconstructions, but not in the DR1+DESY5 one \footnote{For more information on the reconstruction methods see Ref.\cite{Jiang:2024xnu}.}.  While the Chevallier-Polarski-Linder (CPL)  fit, the $w_{0}-w_{a}$ parametrization often considered when interpreting DESI data, is able to partially capture the data characteristic behaviour,  it cannot fit all the  features, since the model does not have sufficient  complexity. Achieving this  may require an oscillatory/non-monotonic behavior of the Hubble expansion rate \cite{Jiang:2024xnu}. More recently, some works  (see \cite{Kessler:2025kju} for instance)  found evidence for an oscillating behaviour of the dark energy equation of state also from the more recent DR2 data.

While a more conclusive result will need to wait for more precise future data, we believe it is now important to further investigate theoretical mechanisms that could account for a non-monotonic dynamical behavior of DE.  As discussed above, the back-reaction effect of super-Hubble fluctuations is one mechanism that could naturally lead to an oscillatory behavior for the dark energy density, and consequently for the Hubble parameter, without the need for extra ingredients beyond the standard $\Lambda$CDM model.  This back-reaction effect is present in any cosmological  model whenever  super-Hubble cosmological fluctuations are present.

In the present work we propose a parametrization to account for the back-reaction effect of super-Hubble fluctuations by means of an oscillating effective energy density component. This may provide a full non-perturbative description of the back-reaction effect. We then analyze such a model considering data from the radial component of the Baryon Acoustic Oscillations from the  recent DESI DR2 data.  We  compute the Hubble parameter at low redshifts as predicted by this model, and compare the result with the data in order to place constraints on the maximum allowed   amplitude of the oscillations. We also analyze the predictions of the model for the effective dark energy equation of state, for the deceleration parameter, and for $H_{0}$. 

Our study can be considered to be a first-step analysis limited to background quantities, aiming only to highlight the importance of not neglecting the back-reaction mechanism. 
Nevertheless, in order to obtain more conclusive results, it is necessary to further analyze the predictions of the model, in particular taking into account the induced cosmological perturbations. This will allow us to test the model using  additional BAO data beyond only the radial component, as well as using further data from other cosmological sources.  This analysis is left for a future work.  

The paper is organized as follows: In Section 2 we propose a new parametrization to describe the back-reaction of super-Hubble fluctuations in the late universe. In Section 3 we discuss the data analysis methods which we use to compare our model predictions with the recent DESI results. In Section 4 we present our results and we conclude in Section 5.

\section{A Model for the Back-Reaction of Super-Hubble fluctuations}
\label{sec2}

The Einstein field equations, as its well known, are highly nonlinear. Even at a  classical level fluctuations at second order affect the background. This effect is called back-reaction. Matter fluctuations induce metric fluctuations through the Einstein equations. These metric fluctuations  also back-react on the homogeneous background metric. The full evolution of the metric is governed by the Einstein equations,
\begin{equation}\label{einstein}
G_{\mu \nu} \, = \, 8\pi G T_{\mu \nu},
\end{equation}
where $G_{\mu \nu}$ is the Einstein tensor of the metric $g_{\mu \nu}$, $G$ is Newton's gravitational constant and $T_{\mu \nu}$ is the energy-momentum tensor of matter.

Metric and matter (taken for simplicity to be a scalar matter field $\varphi$) can be written, at linear order in the amplitude of cosmological fluctuations, as
\begin{equation}\label{metric}
ds^{2} \, = \, (1+2\phi(x,t))dt^{2} - a(t)^{2}[1-2\psi(x,t)\gamma_{ij}dx^{i}dx^{j}],
\end{equation}
and
\begin{equation}\label{matterpert}
\varphi(x,t) \, = \, \varphi_{0} + \delta \varphi(x,t).
\end{equation}
In the absence of anisotropic stress, the scalar metric fluctuation variables $\phi$ and $\psi$ in eq.\ref{metric} are equal. To write the metric, we have chosen a particular coordinate system (longitudinal gauge) \footnote{See e.g. \cite{Mukhanov:1990me} for a review of the theory of cosmological perturbations.}. In Eq. \ref{metric} $\gamma_{ij}$ is the background metric of the constant time hypersurfaces. We assume vanishing spatial curvature and then $\gamma_{ij}=\delta_{ij}$.  In the above equations, the background metric is given by the scale factor $a(t)$ and the background matter by $\varphi_{0}(t)$. We will only consider here the effects of scalar metric fluctuations.

In order to see how the linear fluctuations affect the background, we substitute the ansatz \ref{metric} and \ref{matterpert} into the Einstein equations  and expand to second order. The zero'th order terms satisfy the background equations. The linear terms are assumed to obey the linear perturbation equations.  In this case, the Einstein equations are not satisfied at quadratic order.  Back-reaction is a way to correct for this at quadratic order. The quadratic terms on the left hand side of \ref{einstein} can be moved to the right hand side of the equation, where they form an effective energy-momentum tensor when combined with the quadratic terms in $T_{\mu \nu}$ \cite{Abramo:1997hu, Mukhanov:1996ak},
\begin{equation}\label{tau}
\tau_{\mu \nu} \; \equiv \; < T_{\mu \nu}^{(2)} - \frac{1}{8\pi G} G_{\mu \nu}^{(2)}>,
\end{equation}
where the superscript (2) indicates the order of the terms.

At second order, the linear perturbations modify the background metric.  This effect can be described by introducing a modified background metric 
\begin{equation}
g_{\mu \nu}^{bg}(t) \, \equiv \, g_{\mu \nu}^{0}(t) + \delta g^{bg,2}(t) \, ,
\end{equation}
where the first term is the original background metric and the second reflects the quadratic corrections to the background. 

By taking the spatial average of the Einstein equation \ref{einstein}, expanded to second order, we can describe the corrections to the background. Following the method proposed in \cite{Abramo:1997hu, Mukhanov:1996ak} (and reviewed in \cite{Brandenberger:2002sk}) we can write the resulting equation in  the form,
\begin{equation}
G_{\mu \nu}(g_{\alpha \beta}^{bg}) \, = \, 8\pi G \bigl( T_{\mu \nu}^{(0)} + \tau_{\mu \nu} \bigr) \, ,
\end{equation}
where the first term on the right hand side of the equation is the contribution of the background matter field to $T_{\mu \nu}$. Improved studies of the back-reaction of super-Hubble fluctuations focus on physically measurable quantities such as the spatial average of the local Hubble expansion rate \cite{Geshnizjani:2003cn, Marozzi:2012tp, Finelli:2011cw, Gasperini:2009wp, Gasperini:2009mu, Marozzi:2010qz, Marozzi:2011zb, Abramo:2001dd, Losic:2005vg, Losic:2006ht, Brandenberger:2004ix, Afshordi:2000nr} even beyond perturbation theory \cite{Brandenberger:2018fdd}. Some of these results have pointed to a self adjusting mechanism for the cosmological constant, as first discussed in \cite{Brandenberger:1999su}. 

The origin of the self-adjustment mechanism for the relaxation of the cosmological constant is as follows \cite{Brandenberger:1999su}.: in an accelerating phase, fluctuation modes exit the Hubble radius and freeze out. Cosmological perturbations induced by super-Hubble matter fluctuations contribute to the effective energy-momentum tensor which has the form of a negative cosmological constant (negligible kinetic and tension energies leading to an effective equation of state $w = -1$, and negative effective energy density - negative since matter induces a potential well and the negativity of the gravitational energy overwhelms the positivity of the matter energy for super-Hubble modes). This will induce a decrease in the effective cosmological constant by counteracting the effect of the pre-existing $\Lambda$. But once the energy density $\rho_{DE}(t)$ in the effective cosmological constant has fallen below that of matter, accelerated expansion will stop, modes will re-enter the Hubble radius, the decline in $\rho_{DE}(t)$ halts, and within a fraction of the Hubble time, a renewed buildup of the backreaction effect takes place. 
 
In light of this, we propose here the following parametrization for the effective energy density associated to the back-reaction  of super-Hubble fluctuations:
\begin{equation}
\label{rhoBR}
\rho_{BR} \ = \, A(t) \; cos(\omega t + \phi) \;  \rho_{m}(t),
\end{equation}
where $A$ is a dimensionless amplitude, $\omega$ is the frequency of the oscillation, $\phi$ is a phase and $\rho_{m}$ is the background density of matter. The term $ \omega t$ in the above equation can be written in terms of the redshift $z$ as $\omega t \equiv n  z$,  with $n$ being a constant, which we expect to have an order of magnitude not much smaller or bigger than one, since the typical time scale of oscillations is expected to be set by the Hubble time. This ansatz expresses the idea that, after the time $t_{eq}$ of equal matter and radiation, the back-reaction term should track the energy density $\rho_m$ of matter \footnote{ Note that an oscillatory contribution to the dark energy density was recently also shown to emerge from a model in which the dark energy field emerges from the dynamics of a highly curved field space motivated by string theory \cite{Payeur:2024kyy}. }

A priori it is expected that the amplitude $A(t)$ should be time-dependent. However, in the present work, we will focus on the late time predictions, since we are going to analyze our model in light of data in the range $0\leq z<2.5$. For this small redshift interval we believe it is reasonable to consider $A(t)$ to be constant. Therefore, in the rest of the paper we are going to consider the following expression for the effective density from the back-reaction effect
\begin{equation}
\rho_{BR} \ = \, A \; cos(nz + \phi) \; \rho_{m}(t) \, .
\end{equation}

The contribution of the energy density from the back-reaction effect is added to the total energy of the universe, in such a way that the first Friedmann equation can be written as,
\begin{equation}
\label{friedmann1}
3H^{2} \, = \, 8 \pi G (\rho_{m} + \rho_{BR})+ \Lambda c^{2}, 
\end{equation}
where we have kept a bare cosmological constant term $\Lambda$. We will set the speed of light $c$ to one in the following.

By substituting Eq.\ref{rhoBR} in the above equation, we can write the first Friedmann equation as follows,
\begin{equation}
\label{friedmann2}
H^{2} \, = \, H_{0}^{2} \;\left( (1 + A \; cos(nz + \phi)) \;  \Omega_{m0} \;  (1+z)^{3} + \Omega_{\Lambda 0}\right),
\end{equation}
where \footnote{The subscripts $0$ indicate the quantities at the present time.}
\begin{equation}
\Omega_{m0} \, = \, \rho_{m0}/\rho_{c0}, \; \; \; \; \; \; \;
\Omega_{\Lambda 0} \, = \, \rho_{\Lambda 0}/\rho_{c0},  \; \; \; \; \; \; \; \rho_{c0} \, = \, 3H_{0}^{2}/8\pi G \, .
\end{equation}

Another important quantity to be analyzed is the deceleration parameter, which must show a transition from a decelerated regime to the accelerated one as the dark energy starts dominating. The deceleration parameter is $q \equiv H'/(aH) -1$ where the prime indicates derivative with respect to the redshift. By using  the above equation, we obtain for the deceleration parameter the following expression,
\begin{equation}
\label{deceleration}
q = \frac{(1 + A \; cos(nz + \phi)) \;  \Omega_{m0} \;  3(1+z)^{3} - A sin(nz+\phi)\; n \; \Omega_{m 0}(1+z)^{4}}{2[(1+Acos(nz+\phi))\Omega_{m0}(1+z)^{3}+\Omega_{\Lambda 0}]} -1.
\end{equation}
Concerning the phase $\phi$, we will choose it in a way that the back-reaction mechanism begins at a redshift $z=3400$ with $\rho_{BR}=0$, when it starts building up. This choice is reasonable since we are taking the back-reaction energy density as being proportional to the matter energy density, and before recombination it would rather track the radiation component.

We can now compute the equation of state associated to the back-reaction effective fluid. The continuity equation is not valid in the usual form for this fluid, since there is an energy transfer from the cosmological fluctuations to this effective component \cite{Abramo:1997uy}. In this case, in order to obtain $w_{BR}$, we can consider the second Friedmann equation for the total fluid (considering all the components of the Universe),
\begin{equation}
\label{secondfriedmann}
\dot{H}+H^{2} = - \frac{4\pi G}{3}\left(\rho_{tot} +\frac{3p_{tot}}{c^{2}}\right) =  - \frac{4\pi G}{3}\left(\frac{3H^{2}}{8\pi G} - \frac{3\rho_{\Lambda}}{c^{2}} + \frac{3w_{BR}\; \rho_{BR}}{c^{2}}\right).
\end{equation}
By isolating $w_{BR}$ in the above equation and considering that $\dot{H}=-H' H/a$, we obtain the following expression for the equation of state,
\begin{equation}
\label{eqofstate}
\begin{split}
    w_{BR} &= \left(\frac{H' H}{a} - \frac{3H^{2}}{2} + \frac{3  H_{0}^{2}\; \Omega_{\Lambda 0}}{2}\right)\frac{1}{4\pi G} \; \frac{1}{\rho_{BR}},\\
    &= -\frac{n\left(z+1\right)}{3\tan\left(n(z-3400)\right)}
\end{split}
\end{equation}

In Section IV we will illustrate the behavior of the Hubble parameter at low redshifts, the deceleration parameter and the effective equation of state for chosen values of the frequency and amplitude in Eq.\ref{rhoBR}.  

In this work we will restrict our analysis to the background dynamics. We then do not need to make extra assumptions on  the cosmological model  at perturbative level. A further discussion of the perturbative quantities of the model is left for a future work. While this allow us to make more general predictions concerning the back-reaction mechanism, we consider this study to be a preliminary analysis. In order to obtain more conclusive results, it would be necessary to go beyond the background analysis and test the model with a more complete cosmological  dataset. However the expression above already provide us some insights in light of specific datasets, as for instance the radial component of BAO, which we are going to discuss in the next section. 

\section{Methods and Data Analysis}
\label{sec3} 

The baryon acoustic oscillation (BAO) method is one of the main methods already developed to measure the expansion history of the Universe (for a more complete discussion see for instance \cite{Weinberg:2013agg,Chen:2024tfp, DESI:2024uvr}). In the pre-recombination Universe, acoustic oscillations in the baryon-photon fluid  imprinted a characteristic scale on the clustering of matter, which manifests itself as a bump in the two-point correlation function of matter (or  as a series of oscillations in  the power spectrum). The comoving scale of this feature, is given by the sound horizon at the end of the baryon drag epoch, $r_{d} \approx 100 h^{-1} Mpc$. This quantity depends on the photon and baryon content of the universe at this epoch, and it is well  constrained by CMB  measurements. The BAO feature can be  traced by galaxies, quasars, and the Lyman-$\alpha$ forest. 

The BAO signature can be measured in two  ways: one can use the angular diameter distance to determine the physical length scale corresponding to a certain angle on the sky of the observer, or one can use the difference in redshift  to infer some physical radial distance. Therefore, there are measurements of the apparent size of the BAO standard ruler perpendicular and parallel to the line of sight, allowing one to constrain the angular diameter distance $D_{A}(z)$ and the Hubble parameter $H(z)$, respectively. If the Hubble parameter of our Universe is actually different than the one predicted by the model considered, then there will be differences in the radial BAO scale. 

We can therefore decompose the BAO locations into transverse and line-of-sight dilation parameters, which are denoted by the Alcock-Paczynski-like dilation parameters $\alpha_{\perp}$ and $\alpha_{||}$, which are linked to the angular diameter distance, and to the Hubble parameter, at a given redshift. Here in this work we are interested in the line-of-sight parameter, which is defined as \cite{DESI:2024uvr},
\begin{equation}
    \alpha_{||} \,   \equiv \, \frac{H^{fid}(z) \; r_{d}^{fid}}{H(z) \; r_{d}},
\end{equation}
where the “fid” denotes the quantities measured in the fiducial cosmology, which is the Planck-normalized $\Lambda$CDM model used to convert redshifts into distances in the survey \footnote{Using the notation of Ref.\cite{DESI:2025zgx}, we can write $\alpha_{||}=\alpha_{\text{iso}}\alpha_{\text{AP}}^{2/3}$.}.  In the above, $r_{d}$ is the comoving sound horizon evaluated at the redshift at which baryon-drag optical depth is equal to unity, and it depends on quantities at redshifts higher than that. In the case of the back-reaction model presented in Section II, the quantity  $r_{d}$ will be considered to be  equal to the fiducial one, since we are here not investigating the effects of back-reaction at earlier times.   Therefore in our analysis we are going to consider the approximate expression\footnote{For a similar analysis in a different context, see for instance the work of Ref\cite{Clifton:2024mdy}},
\begin{equation}
    \alpha_{||} \, \approx \, \frac{H^{fid}(z)}{H(z)} \implies \frac{H(z)-H^{fid}(z)}{H^{fid}(z)} = \frac{\Delta H(z)}{H^{fid}(z)} = \frac{1}{\alpha_{||}} - 1,
\end{equation}
where $\Delta H \equiv H(z) - H^{fid}(z)$. We are going to use information from the DESI DR2 galaxy, quasar and Lyman-$\alpha$ results \cite{DESI:2025zgx} to interpret the constraints on $\alpha_{||}$  in terms of constraints on $\Delta H$ at  given redshifts.

The data we are going to consider are described in the following table:
\begin{table}[H]
    \centering
    \begin{tabular}{|l|c|c|c|c|}
        \hline
         Tracer &  Redshift & $\alpha_{||}$ & $\Delta H/H^{fid}$\\
        \hline
        LRG1 & $0.51$ & $0.9615419512\pm 0.01922057331$ & $0.03999622561 \pm  0.0207888212$ \\
         \hline
        LRG2 & $0.706$ & $0.9647661889\pm 0.01642269702$ & $0.03652057001 \pm 0.01764413334$\\
         \hline
        ELG2 & $1.321$ & $0,9932025202\pm 0.01256687344$ & $0.006844001718 \pm 0,01273947749$ \\
         \hline
        QSO  & $1.484$ & $1.02675147\pm 0.02099900114$ & $-0.02605447469 \pm 0.01991902012$\\ 
         \hline
        Lya  & $2.33$ & $1.008212343\pm 0.01700856044$ & $-0.008145449517 \pm 0.01673260419$\\ 
         \hline
        LRG3  & $0.922$ & $0.9953917692\pm 0.04052811979$ & $0.004629564906 \pm 0.04090424355$\\
         \hline
        ELG1  & $0.955$ & $1.001814039\pm 0.01652471677$ & $-0,001810754569 \pm 0,01646492654$\\
   
        \hline
    \end{tabular}
    \caption{\label{tab:1} Dataset used for $\alpha_{||}$  from the DESI DR2 galaxy, quasar and Lyman-$\alpha$ results taken from Ref.\cite{DESI:2025zgx}, and its interpretation in terms of $\Delta H/H^{fid}$ at  given redshifts.}
    \label{tabela_dados}
\end{table}

Our procedure is the following: We first solve Eq.\ref{friedmann2} (with $\Lambda_0$ taken to be the same as in the fiducial cosmology) in order to obtain H(z) predicted by the back-reaction model. From this we can plot the resulting function $\Delta H/H_{\Lambda CDM}$ for the model together with the observational values and error bars shown in Tab.\ref{tabela_dados}. However, in order to plot the quantity $\Delta H/H_{\Lambda CDM}$ predicted by the back-reaction model, the parameters in equation \ref{rhoBR} must be fixed. As mentioned earlier, the phase is fixed in such a way that $\rho_{BR}$ is zero at redshift $z=3.400$. The amplitude $A$ and the "frequency", $n z$, remain to be fixed. We will focus on values of the amplitude which are of the order of $A \sim 10^{-1}$. Larger values of the amplitude would give rise to too large oscillations of the Hubble parameter, while smaller values would not lead to observable differences compared to our fiducial cosmology. We cannot exactly determine the amplitude from the considerations in Section II since this would require a non-perturbative calculation that has not yet been achieved. On the other hand, the considerations in Section II imply that the frequency of oscillation is set by a fraction of the Hubble rate. Hence, we would expect that a non fine-tuned frequency would correspond to a value of $n$ with order of magnitude approximately equal to one.  

We show the results obtained  in the next Section.

\section{Results}
\label{sec4} 

\begin{figure*}[h!]
    \centering
    \includegraphics[width=0.85\textwidth]{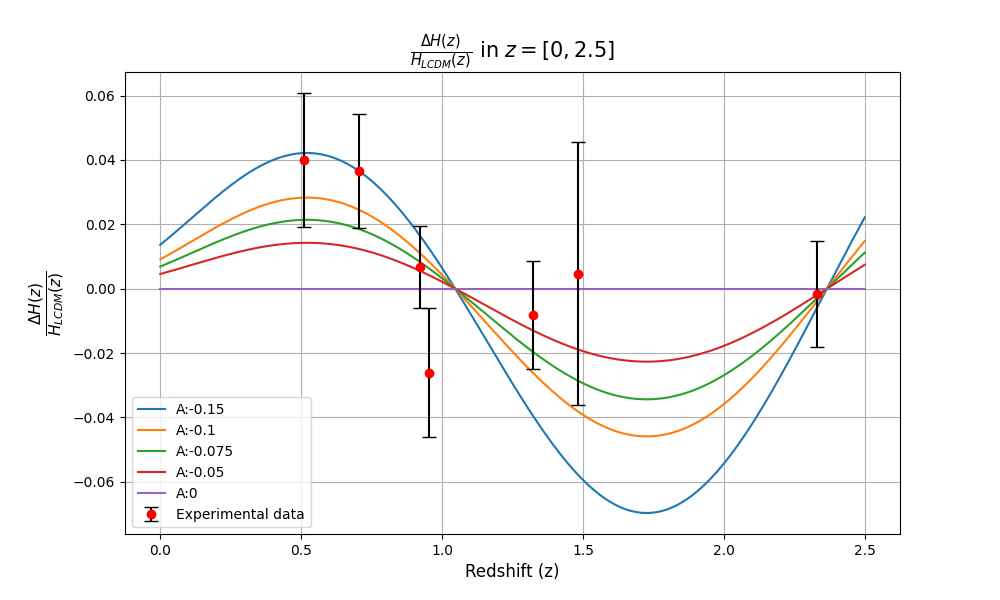}
    \caption{ $\Delta H/H$ predicted by the back-reaction model compared with DESI DR2 data}
    \label{fig:I}
\end{figure*}

In this section we show the predictions of the back-reaction model compared with the data of Ref.\cite{DESI:2025zgx}. In Fig. \ref{fig:I} we show the late-time behavior of the quantity $\Delta H/H_{\Lambda\text{CDM}}$ (curves in color) as predicted by the back-reaction model,  for values of the amplitude ranging from zero (which represents $\Lambda\text{CDM}$) to $A=-0.15$ and for a fixed frequency characterized by the parameter $n\approx 2.4$ introduced in Eq.\ref{friedmann2}. These predictions are compared with the DESI DR2 results (red dots and error bars). Different shapes of the theoretical curve can be obtained using different values of the frequency and amplitude. We choose, just for the sake of illustration, examples of values that provide a shape for the curves which tracks the DESI data better than the standard $\Lambda$CDM model. In addition, in choosing the frequency value to be  illustrated in  the Figures, we selected  models which can alleviate the Hubble tension even slightly, and we disregard  models with too high frequencies ($n \gtrsim 5$) since they could lead to smaller $\chi^2$ simply due to overfitting.  Among  the frequency values that satisfy these criteria, we chose the one that provides the smaller $\chi^2$, which corresponds to $n \approx 2.4$. However, higher values for the frequency can  be considered,  also providing a $\chi^2$ value lower than the $\Lambda$CDM.  Fig.\ref{DeltaH_DR1} in the Appendix shows the same plot as Fig.\ref{fig:I} but for DR1 data. Comparing  Fig.\ref{fig:I} with Fig.\ref{DeltaH_DR1} we see that the evidence of an oscillating dark energy component is stronger in DR2 than in DR1.

Note that the values of the free parameters (amplitude and frequency) which we have used are within the range of values that can be considered as ``reasonable" from the theoretical point of view. In particular, the frequency is in the range expected from qualitative arguments. 

Although the results look promising, our preliminary analysis should not be viewed as conclusive in terms of statistical evidence. With  future data releases, the error bars are expected to shrink, and by combining those with  datasets from other sources it should then be possible to obtain more solid conclusion on the preference of the data for $\Lambda$CDM or a dynamical dark energy. The objective of Fig. \ref{fig:I} is to show the range of amplitudes of the oscillating dark energy term which can be in agreement with the DESI constraints. We can see that, for the chosen value $n \sim 2.4$, values as low as $|A| \sim 0.05$ can fit the data within the observational uncertainties.
 
As a preliminary analysis, we  performed a test minimizing the $\chi^{2}$ of the model with respect to the DESI DR2 data. 
The method used consists of executing a set of minimization routines, such that for each minimization sequence we chose the initial values of $A$ to lie within the range $[-0.1, 0.1]$, and those of $n$ within $[3, 5]$ . These intervals correspond to initial guesses of suitable parameter values and are starting points for the parameter exploration. 

\begin{figure*}[h!]
    \centering
    \centering
    \includegraphics[width=0.85\textwidth]{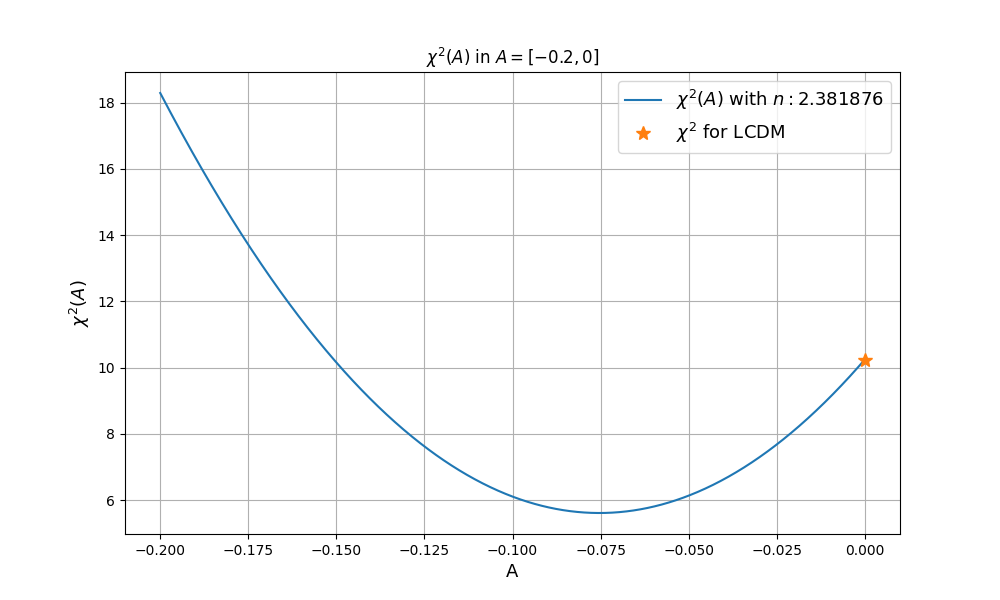}
    \caption{The value of $\chi^2$ for the back-reaction model with respect to DESI data as a function of the parameter  $A$. The value for $\Lambda$CDM is represented by a star.}
    \label{fig:chi}
\end{figure*}

The expression to be minimized is the $\chi^2$ of $\Delta H(z)/ H_{\Lambda CDM}$ (which expresses the relative difference between the proposed model and the standard cosmological model as a function of redshift) with respect to the experimental data from DESI DR2. As described previously, the DESI experimental results were interpreted in terms of $\Delta H$, and the error propagation was calculated.  Some minimization results lead to a shape more in line with the one indicated by DESI data than others. However, in practically all the cases considered, for the mentioned initial values, the $\chi^2$ obtained 
were smaller than the one of the $\Lambda$CDM model. On the other hand, these analysis were restricted to amplitudes with order of magnitude $A=\mathcal{O}(10^{-1})$ or smaller.  For larger values of $A$ the fit is poor. This is shown in Fig.\ref{fig:chi}, where  the value of $\chi^{2}$ is plotted as a function of the amplitude.  
From this same figure we can also see that for values of the amplitude higher than zero and lower than $A \sim-0.15$ the back-reaction model provides a fit worse than the standard model (which is recovered for the value $A=0$, the star in the  plot). We can also see from Fig. \ref{fig:chi} that the minimum value of the $\chi^2$ is shifted with respect to the $\Lambda$CDM value. A similar plot using the data from DR1 can be seen in the Appendix. By comparing the $\chi^2$ as a function of the amplitude for the two datasets, DR1 and DR2, we see that the minimum value of the $\chi^2$ does not change considerably. However from the overall analysis, we obtain that the frequencies that tend to provided  best fitting for the DR2 data are in general smaller then the ones that fits better  the DR1 data. This general tendency seems to be in agreement with the results of Ref. \cite{Ormondroyd:2025iaf}. While the error bars still do not allow us to conclude that our model performs better than $\Lambda$CDM, these results can be viewed as an indication of the importance of further studies of this model, rather than a firm statistical conclusion.

\begin{figure*}[h!]
    \centering
    \includegraphics[width=0.85\textwidth]{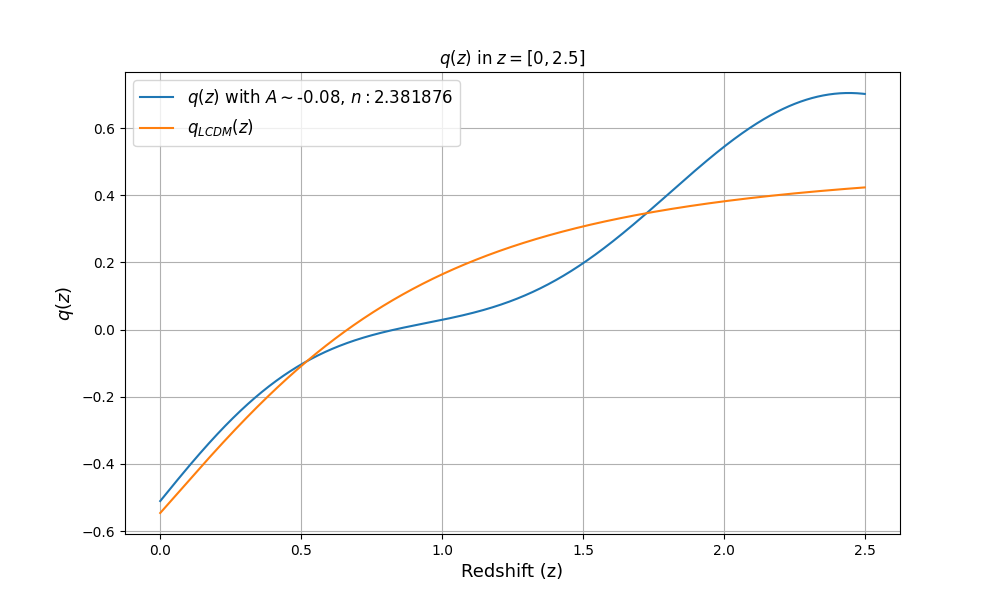}
    \caption{Deceleration parameter from $z=2.5$ to $z=0$ predicted by the back-reaction model, together with the $\Lambda$CDM prediction.}
    \label{fig:III}
\end{figure*}

\begin{figure*}[h!]
    \centering
    \begin{subfigure}{0.49\textwidth}
        \centering
        \includegraphics[width=\textwidth]{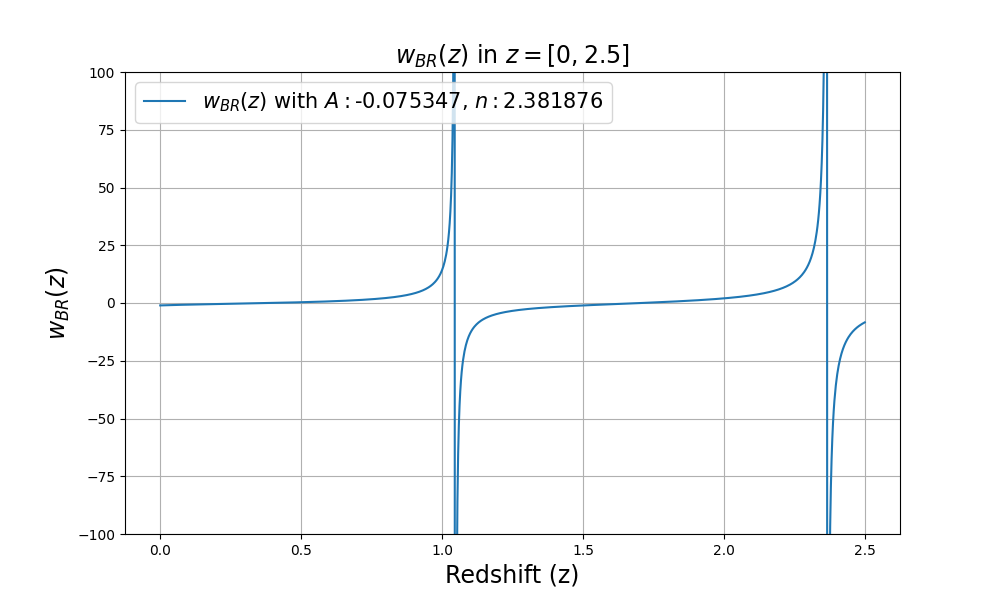}
        \label{fig:IV_a}
    \end{subfigure}
    \hfill
    \begin{subfigure}{0.49\textwidth}
        \centering
        \includegraphics[width=\textwidth]{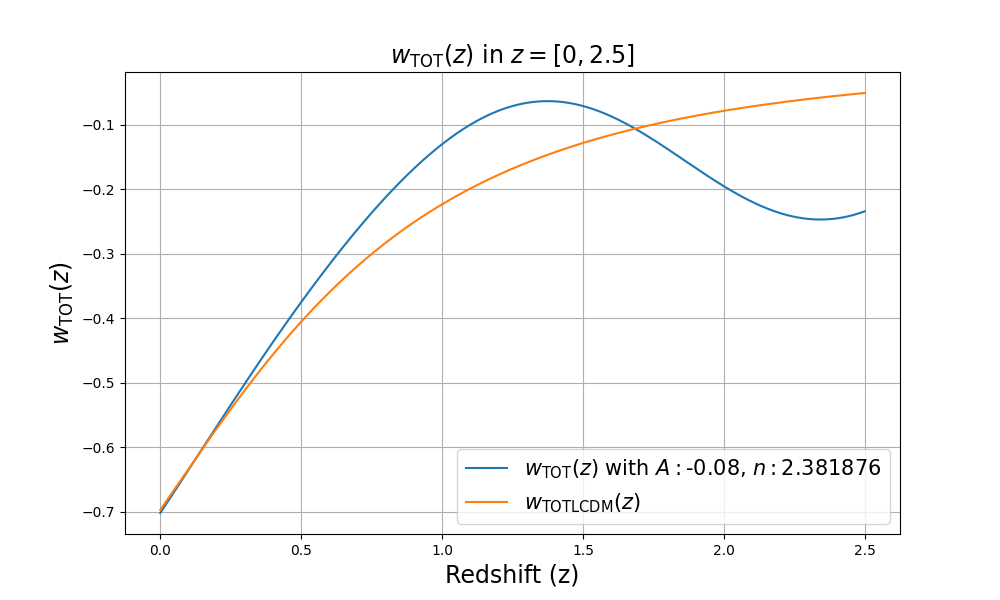}
        \label{fig:IV_b}
    \end{subfigure}
    \caption{The evolution of the effective equation of state of the back-reaction fluid (left) and the total effective equation of state (right).}
    \label{fig:comparison}
\end{figure*}

In Figs \ref{fig:III} and \ref{fig:comparison} we plot, respectively, the deceleration parameter and the equations of state (EOS) of the back-reaction model (the EOS for the effective backreaction fluid and the total EOS including all the components), according to the equations in Section II. The pair of values $(A \sim -0.08, n \sim 2.4)$ was chosen as an illustrative example in all these figures. We can see in Fig.\ref{fig:III} that this model is able to correctly predict the transition from decelerated to accelerated expansion around the expected redshift.

In Fig.\ref{fig:comparison} (left) we illustrate  the evolution of the equation of state for the effective back-reaction fluid,  given by Eq.\ref{eqofstate}. We considered again the same values for the parameters $A$ and $n$. We can see in this plot that the equation of state shows the average value $w \sim 0$. One can also see that, at the exact redshifts when $\rho_{BR}$ crosses zero, a spike can be seen in the plot of $w_{BR}$. This should be expected from the fluid equation of state $w_{BR}\equiv p_{BR}/\rho_{BR}$. The spikes correspond to the points where the quantity $\Delta H/H_{ \Lambda CDM}$ crosses zero in Fig.\ref{fig:I}. However, this behavior does not correspond to a problem \footnote{A similar behavior can be seen for instance, in the model analyzed in Ref.\cite{Tiwari:2024gzo}.} since the observable quantities in cosmology correspond to integrals of the equation of state over a period of time. Therefore these spikes are not expected to be present in any cosmological observable. In addition, the contribution from the back-reaction effective fluid is always subdominant compared to the other components of the Universe. The effective equation of state of the total fluid has the expected overall behavior, as shown in Fig.\ref{fig:comparison} (right), i.e., it oscillates around the $\Lambda CDM$ prediction.  

\begin{figure*}[h!]
    \centering
    
    \centering
    \includegraphics[width=0.85\textwidth]{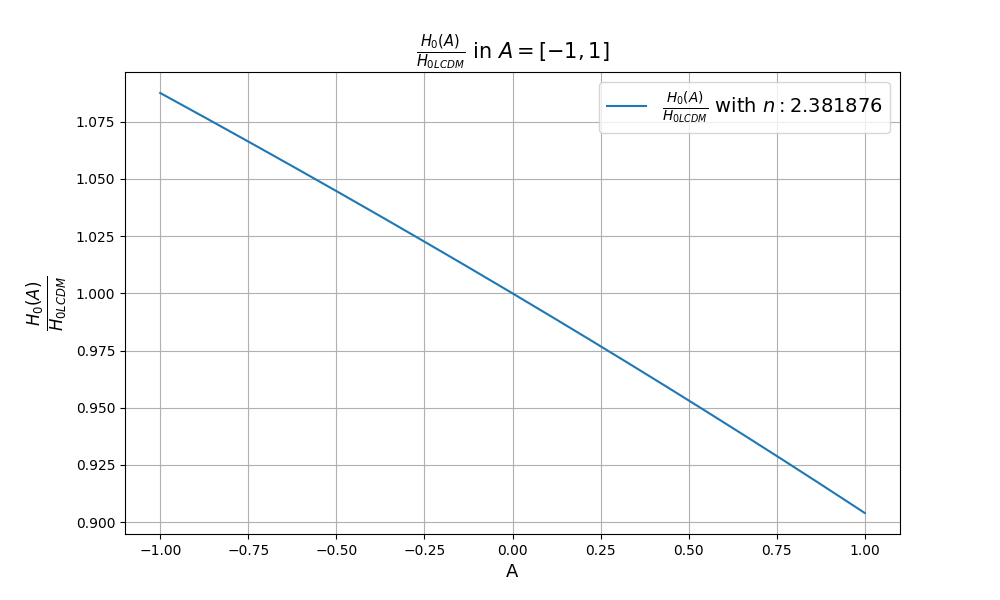}
    
    \caption{$H_0$ as a function of $A$, for $n$ chosen from the minimization method performed.}
    \label{fig:V}
\end{figure*}

Another major challenge that has become more pressing following the DESI data is the Hubble tension problem \cite{Colgain:2025nzf, Poulin:2025nfb, Pang:2025lvh}.
Recent results from DR2 bolster the case for a dynamical dark energy scenario exhibiting phantom crossing behavior. This framework yields a lower inferred value of the Hubble constant, thereby intensifying the Hubble tension — which can be regarded as a symptom of potential underlying inconsistencies in the standard model.  Therefore it is interesting to analyse in which cases, for which values of the parameters, the back-reaction model considered  can provide some relief to the Hubble tension. In order to shed some light to this question we perform an analysis of the value of $H_0$  predicted for different combinations of the parameters $n$ and $A$. The analysis uses the  method described previously. We show in Fig. \ref{fig:V} a plot of $H_0$ ($H_0 = H_0(A, n_{\text{best}})$) as a function of $A$, where $n_{\text{best}}\sim 2.4$ is the value of $n$  chosen from the minimization method performed. 
We can see in the plot that, for the representative chosen values of the parameters, the $H_0$ predicted by our model is very close to the $\Lambda$CDM one. 
 This is partly due to the choice of the fixed A and n and also due to the fact that the back-reaction contribution is proportional to $\rho_m$, therefore decreasing at later times, causing the difference of our model with respect to $\Lambda$CDM also to decrease.

From all the results of our minimizations, the highest value that was obtained was $H_{0}\approx1.009 \; H_{0\Lambda CDM}$, which is achieved for $A\approx-0.07$ and $n\approx2.06$. For the  amplitude $A= - 0.05$, for instance, we obtain $H_{0}/H_{0\Lambda CDM} = 1 \pm 10^{-3}$, i.e., it's basically indistinguishable from $\Lambda CDM$. Nevertheless, it could be possible to solve the Hubble tension given a sufficiently high absolute values of the amplitude. However, from Fig \ref{fig:I}  we can see that such high absolute values of $A$ do not provide a good fit to DESI data.   If we extrapolate and investigate the Hubble constant value for an amplitude as low as $A=-1$, we obtain  $H_{0} \approx 1.09 \; H_{0\Lambda CDM}$,  considering  $n=n_{\text{best}}$. 

It is important to emphasize again that a priori there is no reason that the amplitude $A$ should be independent of time.  In fact,  from the considerations of back-reaction it is not unreasonable to expect that the effective amplitude of the oscillating contribution to DE performs damped oscillations. In this case,  $A$ could be large at early times and then slowly decay to a value consistent with the current late time data.  This extended scenario might therefore be able to provide a solution of the Hubble tension. On the other hand, without a careful analysis of the fluctuations produced in the model, it is not possible to gauge whether such scenario would be consistent with other constraints, e.g. the value of $\sigma_8$.

While conclusive results concerning the viability of the model will need to wait for a more complete analysis with the future data,  our results shows that the inclusion of  back-reaction effects from super-Hubble fluctuations  can offer valuable insights in light of recent observational data. 

\section{Conclusion}
\label{sec5}

In the present work, we studied a toy model of oscillating dark energy motivated by considerations of the back-reaction of super-Hubble fluctuations in the late universe. We proposed a parametrization of the oscillating dark energy component to account for the back-reaction effect on the background cosmology (based on the conjectured dynamical scaling fixed point solution describing the back-reaction effect at a fully non-perturbative level). We computed the Hubble parameter at low redshifts predicted by the model, obtaining an oscillatory dynamics. We also compared the resulting background cosmology with data from the radial component of the Baryon Acoustic Oscillations obtained from the recent DESI survey \cite{DESI:2025zgx}. We found that, for a frequency scale expected from theoretical considerations, a fit to the recent observations can be obtained for a certain range of the amplitude of the oscillation, which is a free parameter in the analysis, and for a frequency of oscillation motivated by the theory. At the present time the observational error bars are still too large to observe a statistically significant advantage of our model over the standard $\Lambda$CDM cosmology. But, with upcoming observations the error bars will shrink and may lead to a way to extract a non-vanishing amplitude of an oscillating component to dark energy in a reliable way.

We also investigated the impact of our model on the Hubble tension. We found that for a large amplitude $A$ of the oscillatory component (such that the energy density in the oscillating term is comparable to the matter energy density) a sufficiently large change in $H_0$ results to explain the Hubble tension. However such a high amplitude does not provide a good fit to the data. If, on the other hand, $A$ depended on time and decreased from a large value near the time of recombination to a small value at the present time, the Hubble constant tension might be successfully addressable. However, detailed studies of the effects of our model on cosmological fluctuations is required to check the consistency of such a scenario.

The next step should in fact be to analyze the perturbations in our scenario. This would allow us to make comparisons with other observable quantities, e.g. CMB anisotropies, and in particular the Integrated Sachs Wolfe (ISW) effect.  In this context, in Ref.\cite{Pace:2011kb} the behavior of several cosmological observables, such as the linear growth factor, the ISW effect, the number counts of massive structures, and the matter and cosmic shear power spectra were estimated for several oscillating dark energy models. It was shown that, for several of the models considered,  independently of the amplitude and the frequency of the dark energy oscillations, none of these observables showed a significant oscillating behavior as a function of redshift. This is a consequence of the fact that such observables are integrals over functions of the expansion rate along the cosmic history, consequently smoothing oscillatory features below the level of detectability. This analysis was performed under certain assumptions, for instance, supposing that dark energy does not cluster on the scales of interest. In addition, the models studied in that reference have a different behavior compared to the back-reaction model we consider in our work. A complete analysis specific for our scenario is hence still missing, but we believe it might be possible that the main conclusions of the work of \cite{Pace:2011kb} might also hold for our model. As the observables mentioned are integrals over  functions of the expansion rate along the cosmic history, we expect that also in our scenario the oscillatory features will be smoothed, keeping the ISW effect in agreement with current observations. 

Another quantity that is going to be investigated in upcoming work is the sound speed predicted in our scenario. Although it may not necessarily be true that the back-reaction model must reproduce the $\Lambda$CDM behavior for this quantity, one could expect similar predictions in this regard. Since our model accounts only for the back-reaction of super-Hubble fluctuations, one could expect that a quantity related to the dynamics of sub-Hubble fluctuations, as the sound speed, could  follow a similar behavior as the standard model. A further analysis of these aspects will be addressed in a future work.

Our work can be considered to be a first-step analysis limited to the background predictions of the model. 
 Our results, although preliminary, highlights the importance of considering the back-reaction effect of super-Hubble fluctuations in the late Universe, as such back-reaction effects are expected to be present in any realistic scenario. We have shown that this provides a dynamics which is interesting in light of recent observational results.

\appendix

\section{Previous DESI DR1 Analysis}

In this Appendix we show the analysis of the back-reaction model in light of the previous DR1 results for the sake of comparison.
 Using the DR1 data we produce the Table below  with the values for $\alpha_{||}$  from Refs. \cite{DESI:2024uvr} \cite{DESI:2024lzq}, and its interpretation in terms of $\Delta H/H^{fid}$ at given redshifts.

\begin{table}[H]
    \centering
    \begin{tabular}{|l|c|c|c|c|}
        \hline
         Tracer &  Redshift & $\alpha_{||}$ & $\Delta H/H^{fid}$\\
        \hline
        LRG 1 & $0.4-0.6$ & $0.9219 \pm 0.0269$ & $0.0847 \pm  0.0317$ \\
         \hline
        LRG 2  & $0.6-0.8$ & $0.9955 \pm 0.0296$ & $0.0045 \pm 0.0299$\\
         \hline
        LRG 3  & $0.8-1.1$ & $1.0057 \pm 0.0214$ & $-0.0057 \pm 0.0212$ \\
         \hline
        ELG 2  & $1.1-1.6$ & $0.9797 \pm 0.0297$ & $0.0207 \pm 0.0309$\\ 
         \hline
         Lyman-$\alpha$  & $ \sim 2.33$ & $0.993^{+0.029}_{-0.032}$ & $0.007 ^{+0.029}_{-0.032}$\\
   
        \hline
    \end{tabular}
    \caption{\label{tab:1_DR1} Dataset used for $\alpha_{||}$  from the DESI first-year galaxy and quasar results \cite{DESI:2024uvr}, and the Lyman-$\alpha$ forest \cite{DESI:2024lzq}, and its interpretation in terms of $\Delta H/H^{fid}$ at given redshifts.}
\end{table}

\begin{figure*}[h!]
    \centering
    \includegraphics[width=0.85\textwidth]{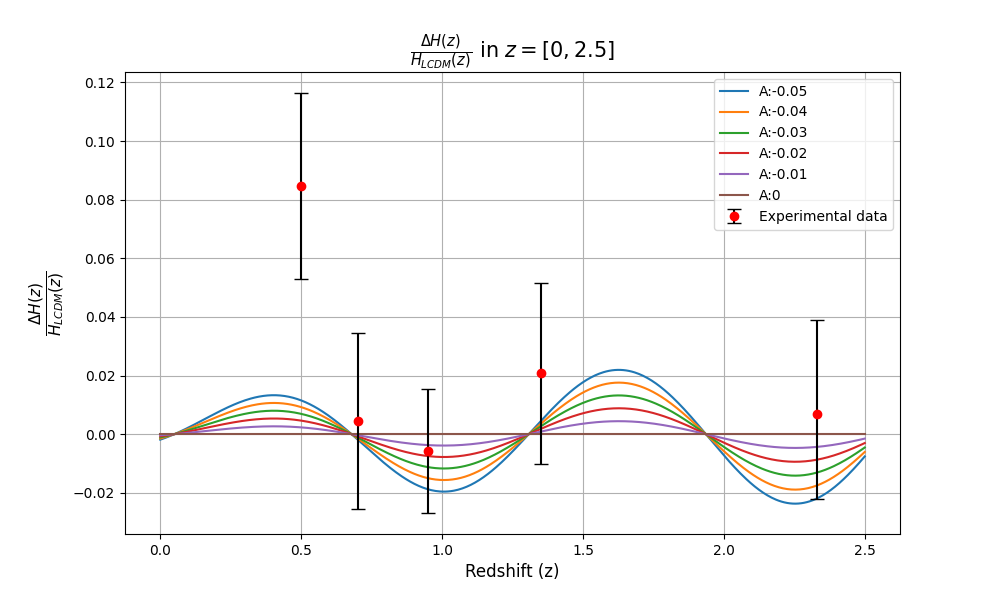}
    \caption{ $\Delta H/H$ predicted by the back-reaction model compared with DR1 data. In all the curves we considered the value $n\sim 5$.}
    \label{DeltaH_DR1}
\end{figure*}

In Fig.\ref{DeltaH_DR1} we
show the late-time behavior of the function $\Delta H/H_{\Lambda CDM}$ (curves in color) as predicted by the back-reaction
model, for values of the parameter A ranging
from zero to $A=-0.15$ and for a fixed value of $n \sim 5$. These predictions are compared  with the DESI DR1 results (red dots and
error bars). Different shapes of the theoretical curve can be obtained using different values of the frequency and
amplitude. To illustrate the method, we select particular sets of parameter values that result in curve shapes that more closely align with the DESI DR1 data.

\begin{figure*}[!t]
    \centering
    \centering
    \includegraphics[width=0.85\textwidth]{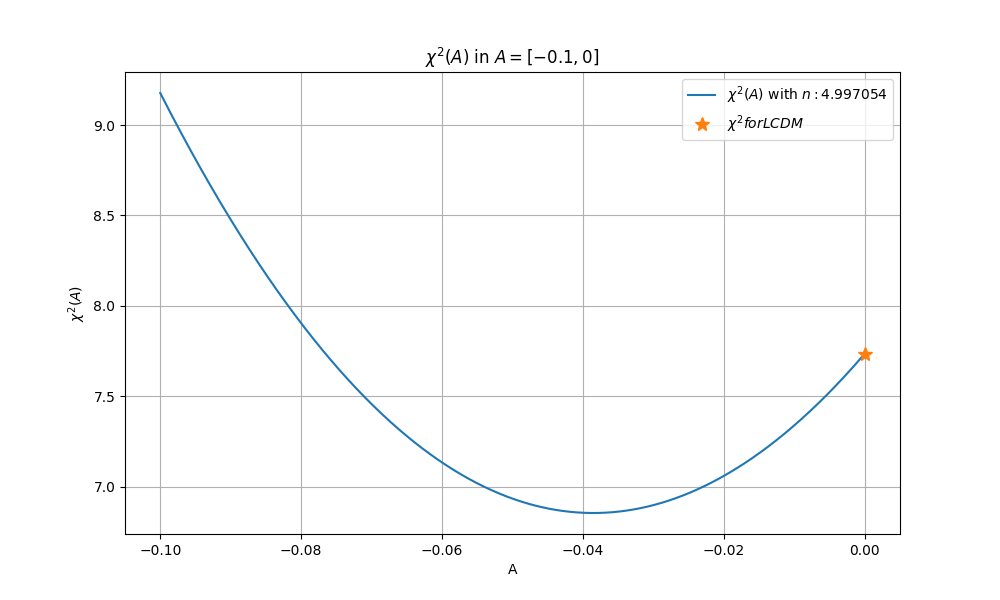}
    \caption{The value of $\chi^2$ for the back-reaction model with respect to DR1 data as a function of the parameter  $A$. The value for $\Lambda$CDM is represented by a star.}
    \label{fig:chi_DR1}
\end{figure*}

We performed a test minimizing the $\chi^2$ of the model with respect to the DR1 data using the same method described in Section IV. In Fig.\ref{fig:chi_DR1}, we show the value of
$\chi^2$ plotted as a function of the amplitude. As in the analysis for DR2 shown in Fig.\ref{fig:I}, we can see that also in this case  higher values
of the amplitude the back-reaction model provides a fit worse than the standard model. Again  we can  see 
that the minimum value of the $\chi^2$ is shifted with respect to the $\Lambda$CDM value (which is recovered for the value A = 0). The minimum is achieved for an amplitude slightly smaller in modulus compared to the case with DR2 shown in Fig.\ref{fig:chi}.

Although the best-fit frequency and amplitude of our model show slight variations when using DESI DR2 instead of DR1, and a modest improvement in the $\chi^2$ value is observed relative to the $\Lambda$CDM model when using DR2, the overall conclusions remain consistent across both DESI data releases.

\FloatBarrier
 
\begin{acknowledgments}
 
We wish to thank Elisa Ferreira for extensive discussions, and E. Colgain for important feedback on the first draft of this paper. M.A.C.A. is supported by Coordenacao de Aperfeicoamento de Pessoal de Nivel Superior (CAPES). L.L.G
is supported by research grants from Conselho Nacional
de Desenvolvimento Cientıfico e Tecnologico (CNPq),
Grant No. 307636/2023-2 and from the Fundacao Carlos
Chagas Filho de Amparo a Pesquisa do Estado do Rio
de Janeiro (FAPERJ), Grant No. E-26/204.598/2024.
 RB is supported in part by NSERC and the CRC program.

\end{acknowledgments}

\bibliographystyle{unsrt}
\bibliography{Bibilography.bib}

\begin{thebibliography}{10}

\bibitem{Polyakov:2012uc}
A.~M. Polyakov.
\newblock {Infrared instability of the de Sitter space}.
\newblock 9 2012.

\bibitem{Polyakov:2007mm}
A.~M. Polyakov.
\newblock {De Sitter space and eternity}.
\newblock {\em Nucl. Phys. B}, 797:199--217, 2008.

\bibitem{Mazur:1986et}
Pawel Mazur and Emil Mottola.
\newblock {Spontaneous Breaking of De Sitter Symmetry by Radiative Effects}.
\newblock {\em Nucl. Phys. B}, 278:694--720, 1986.

\bibitem{Mottola:1985qt}
E.~Mottola.
\newblock {THERMODYNAMIC INSTABILITY OF DE SITTER SPACE}.
\newblock {\em Phys. Rev. D}, 33:1616--1621, 1986.

\bibitem{Mottola:1984ar}
E.~Mottola.
\newblock {Particle Creation in de Sitter Space}.
\newblock {\em Phys. Rev. D}, 31:754, 1985.

\bibitem{Tsamis:1994ca}
N.~C. Tsamis and R.~P. Woodard.
\newblock {Strong infrared effects in quantum gravity}.
\newblock {\em Annals Phys.}, 238:1--82, 1995.

\bibitem{Obied:2018sgi}
Georges Obied, Hirosi Ooguri, Lev Spodyneiko, and Cumrun Vafa.
\newblock {De Sitter Space and the Swampland}.
\newblock 6 2018.

\bibitem{Dvali:2014gua}
Gia Dvali and Cesar Gomez.
\newblock {Quantum Exclusion of Positive Cosmological Constant?}
\newblock {\em Annalen Phys.}, 528:68--73, 2016.

\bibitem{Danielsson:2018ztv}
Ulf~H. Danielsson and Thomas Van~Riet.
\newblock {What if string theory has no de Sitter vacua?}
\newblock {\em Int. J. Mod. Phys. D}, 27(12):1830007, 2018.

\bibitem{Dasgupta:2018rtp}
Keshav Dasgupta, Maxim Emelin, Evan McDonough, and Radu Tatar.
\newblock {Quantum Corrections and the de Sitter Swampland Conjecture}.
\newblock {\em JHEP}, 01:145, 2019.

\bibitem{Bedroya:2019snp}
Alek Bedroya and Cumrun Vafa.
\newblock {Trans-Planckian Censorship and the Swampland}.
\newblock {\em JHEP}, 09:123, 2020.

\bibitem{Bedroya:2019tba}
Alek Bedroya, Robert Brandenberger, Marilena Loverde, and Cumrun Vafa.
\newblock {Trans-Planckian Censorship and Inflationary Cosmology}.
\newblock {\em Phys. Rev. D}, 101(10):103502, 2020.

\bibitem{Brandenberger:2021pzy}
Robert Brandenberger.
\newblock {Trans-Planckian Censorship Conjecture and Early Universe Cosmology}.
\newblock {\em LHEP}, 2021:198, 2021.

\bibitem{Tsamis:1992sx}
N.~C. Tsamis and R.~P. Woodard.
\newblock {Relaxing the cosmological constant}.
\newblock {\em Phys. Lett. B}, 301:351--357, 1993.

\bibitem{Tsamis:1996qq}
N.~C. Tsamis and R.~P. Woodard.
\newblock {Quantum gravity slows inflation}.
\newblock {\em Nucl. Phys. B}, 474:235--248, 1996.

\bibitem{Abramo:1997hu}
L.~Raul~W. Abramo, Robert~H. Brandenberger, and Viatcheslav~F. Mukhanov.
\newblock {The Energy - momentum tensor for cosmological perturbations}.
\newblock {\em Phys. Rev. D}, 56:3248--3257, 1997.

\bibitem{Finelli:2001bn}
F.~Finelli, G.~Marozzi, G.~P. Vacca, and Giovanni Venturi.
\newblock {Energy momentum tensor of field fluctuations in massive chaotic inflation}.
\newblock {\em Phys. Rev. D}, 65:103521, 2002.

\bibitem{Finelli:2003bp}
F.~Finelli, G.~Marozzi, G.~P. Vacca, and Giovanni Venturi.
\newblock {Energy momentum tensor of cosmological fluctuations during inflation}.
\newblock {\em Phys. Rev. D}, 69:123508, 2004.

\bibitem{Marozzi:2006ky}
G.~Marozzi.
\newblock {Back-reaction of Cosmological Fluctuations during Power-Law Inflation}.
\newblock {\em Phys. Rev. D}, 76:043504, 2007.

\bibitem{Brandenberger:1999su}
Robert~H. Brandenberger.
\newblock {Back reaction of cosmological perturbations}.
\newblock In {\em {3rd International Conference on Particle Physics and the Early Universe}}, pages 198--206, 2000.

\bibitem{Ahmed:2002mj}
Maqbool Ahmed, Scott Dodelson, Patrick~B. Greene, and Rafael Sorkin.
\newblock {Everpresent $\Lambda$}.
\newblock {\em Phys. Rev. D}, 69:103523, 2004.

\bibitem{Das:2023hbw}
Santanu Das, Arad Nasiri, and Yasaman~K. Yazdi.
\newblock {Aspects of Everpresent \ensuremath{\Lambda}. Part I. A~fluctuating cosmological constant from spacetime discreteness}.
\newblock {\em JCAP}, 10:047, 2023.

\bibitem{Das:2023rvg}
Santanu Das, Arad Nasiri, and Yasaman~K. Yazdi.
\newblock {Aspects of Everpresent $\Lambda$ (II): Cosmological Tests of Current Models}.
\newblock 7 2023.

\bibitem{Unruh:1998ic}
W.~Unruh.
\newblock {Cosmological long wavelength perturbations}.
\newblock 2 1998.

\bibitem{Geshnizjani:2002wp}
Ghazal Geshnizjani and Robert Brandenberger.
\newblock {Back reaction and local cosmological expansion rate}.
\newblock {\em Phys. Rev. D}, 66:123507, 2002.

\bibitem{Abramo:2001dc}
L.~R. Abramo and R.~P. Woodard.
\newblock {No one loop back reaction in chaotic inflation}.
\newblock {\em Phys. Rev. D}, 65:063515, 2002.

\bibitem{Geshnizjani:2003cn}
Ghazal Geshnizjani and Robert Brandenberger.
\newblock {Back reaction of perturbations in two scalar field inflationary models}.
\newblock {\em JCAP}, 04:006, 2005.

\bibitem{Marozzi:2012tp}
Giovanni Marozzi, Gian~Paolo Vacca, and Robert~H. Brandenberger.
\newblock {Cosmological Backreaction for a Test Field Observer in a Chaotic Inflationary Model}.
\newblock {\em JCAP}, 02:027, 2013.

\bibitem{Finelli:2011cw}
F.~Finelli, G.~Marozzi, G.~P. Vacca, and G.~Venturi.
\newblock {Backreaction during inflation: A Physical gauge invariant formulation}.
\newblock {\em Phys. Rev. Lett.}, 106:121304, 2011.

\bibitem{Gasperini:2009wp}
M.~Gasperini, G.~Marozzi, and G.~Veneziano.
\newblock {Gauge invariant averages for the cosmological backreaction}.
\newblock {\em JCAP}, 03:011, 2009.

\bibitem{Gasperini:2009mu}
M.~Gasperini, G.~Marozzi, and G.~Veneziano.
\newblock {A Covariant and gauge invariant formulation of the cosmological 'backreaction'}.
\newblock {\em JCAP}, 02:009, 2010.

\bibitem{Marozzi:2010qz}
G.~Marozzi.
\newblock {The cosmological backreaction: gauge (in)dependence, observers and scalars}.
\newblock {\em JCAP}, 01:012, 2011.

\bibitem{Marozzi:2011zb}
Giovanni Marozzi and Gian~Paolo Vacca.
\newblock {Isotropic Observers and the Inflationary Backreaction Problem}.
\newblock {\em Class. Quant. Grav.}, 29:115007, 2012.

\bibitem{Abramo:2001dd}
L.~R. Abramo and R.~P. Woodard.
\newblock {Back reaction is for real}.
\newblock {\em Phys. Rev. D}, 65:063516, 2002.

\bibitem{Losic:2005vg}
B.~Losic and W.~G. Unruh.
\newblock {Long-wavelength metric backreactions in slow-roll inflation}.
\newblock {\em Phys. Rev. D}, 72:123510, 2005.

\bibitem{Losic:2006ht}
B.~Losic and W.~G. Unruh.
\newblock {On leading order gravitational backreactions in de Sitter spacetime}.
\newblock {\em Phys. Rev. D}, 74:023511, 2006.

\bibitem{Brandenberger:2004ix}
Robert~H. Brandenberger and C.~S. Lam.
\newblock {Back-reaction of cosmological perturbations in the infinite wavelength approximation}.
\newblock 7 2004.

\bibitem{Afshordi:2000nr}
Niayesh Afshordi and Robert~H. Brandenberger.
\newblock {Super Hubble nonlinear perturbations during inflation}.
\newblock {\em Phys. Rev. D}, 63:123505, 2001.

\bibitem{Comeau:2023gxk}
Vincent Comeau.
\newblock {Perturbative Correction to the Average Expansion Rate of Spacetimes with Perfect Fluids}.
\newblock 4 2023.

\bibitem{Comeau:2023euf}
Vincent Comeau and Robert Brandenberger.
\newblock {Back-reaction of long-wavelength cosmological fluctuations as measured by a clock field}.
\newblock {\em Eur. Phys. J. C}, 84(3):272, 2024.

\bibitem{Giani:2024nnv}
Leonardo Giani, Rodrigo Von~Marttens, and Ryan Camilleri.
\newblock {A novel approach to cosmological non-linearities as an effective fluid}.
\newblock arxiv:2410.15295 2024.

\bibitem{Kolb:2009rp}
Edward~W. Kolb, Valerio Marra, and Sabino Matarrese.
\newblock {Cosmological background solutions and cosmological backreactions}.
\newblock {\em Gen. Rel. Grav.}, 42:1399--1412, 2010.

\bibitem{Kolb:2011zz}
Edward~W. Kolb.
\newblock {Backreaction of inhomogeneities can mimic dark energy}.
\newblock {\em Class. Quant. Grav.}, 28:164009, 2011.

\bibitem{Marra:2011ct}
Valerio Marra and Alessio Notari.
\newblock {Observational constraints on inhomogeneous cosmological models without dark energy}.
\newblock {\em Class. Quant. Grav.}, 28:164004, 2011.

\bibitem{Brandenberger:2018fdd}
Robert Brandenberger, Leila~L. Graef, Giovanni Marozzi, and Gian~Paolo Vacca.
\newblock {Backreaction of super-Hubble cosmological perturbations beyond perturbation theory}.
\newblock {\em Phys. Rev. D}, 98(10):103523, 2018.

\bibitem{DESI:2024aqx}
R.~Calderon et~al.
\newblock {DESI 2024: Reconstructing Dark Energy using Crossing Statistics with DESI DR1 BAO data}.
\newblock 5 2024.

\bibitem{DESI:2024kob}
K.~Lodha et~al.
\newblock {DESI 2024: Constraints on Physics-Focused Aspects of Dark Energy using DESI DR1 BAO Data}.
\newblock 5 2024.

\bibitem{DESI:2024mwx}
A.~G. Adame et~al.
\newblock {DESI 2024 VI: Cosmological Constraints from the Measurements of Baryon Acoustic Oscillations}.
\newblock 4 2024.

\bibitem{DESI:2025zgx}
M.~Abdul~Karim et~al.
\newblock {DESI DR2 Results II: Measurements of Baryon Acoustic Oscillations and Cosmological Constraints}.
\newblock 3 2025.

\bibitem{DESI:2025fii}
K.~Lodha et~al.
\newblock {Extended Dark Energy analysis using DESI DR2 BAO measurements}.
\newblock 3 2025.

\bibitem{Payeur:2024dnq}
Guillaume Payeur, Evan McDonough, and Robert Brandenberger.
\newblock {Do Observations Prefer Thawing Quintessence?}
\newblock 11 2024.

\bibitem{DESI:2025wyn}
Gan Gu et~al.
\newblock {Dynamical Dark Energy in light of the DESI DR2 Baryonic Acoustic Oscillations Measurements}.
\newblock 4 2025.

\bibitem{Brandenberger:2025hof}
Robert Brandenberger.
\newblock {Why the DESI Results Should Not Be A Surprise}.
\newblock 3 2025.

\bibitem{Jiang:2024xnu}
Jun-Qian Jiang, Davide Pedrotti, Simony~Santos da~Costa, and Sunny Vagnozzi.
\newblock {Non-parametric late-time expansion history reconstruction and implications for the Hubble tension in light of DESI}.
\newblock 8 2024.

\bibitem{Escamilla:2024fzq}
Luis~A. Escamilla, Supriya Pan, Eleonora Di~Valentino, Andronikos Paliathanasis, Jos\'e~Alberto V\'azquez, and Weiqiang Yang.
\newblock {Testing an oscillatory behavior of dark energy}.
\newblock {\em Phys. Rev. D}, 111(2):023531, 2025.

\bibitem{Akarsu:2022lhx}
Ozgur Akarsu, Eoin~O. Colgain, Emre \"Ozulker, Somyadip Thakur, and Lu~Yin.
\newblock {Inevitable manifestation of wiggles in the expansion of the late Universe}.
\newblock {\em Phys. Rev. D}, 107(12):123526, 2023.

\bibitem{Colgain:2024mtg}
Eoin~\'O. Colg\'ain and M.~M. Sheikh-Jabbari.
\newblock {DESI and SNe: Dynamical Dark Energy, $\Omega_m$ Tension or Systematics?}
\newblock 12 2024.

\bibitem{Zhao:2017cud}
Gong-Bo Zhao et~al.
\newblock {Dynamical dark energy in light of the latest observations}.
\newblock {\em Nature Astron.}, 1(9):627--632, 2017.

\bibitem{Tamayo:2019gqj}
David Tamayo and J.~Alberto Vazquez.
\newblock {Fourier-series expansion of the dark-energy equation of state}.
\newblock {\em Mon. Not. Roy. Astron. Soc.}, 487(1):729--736, 2019.

\bibitem{Rubano:2003er}
C.~Rubano, Paolo Scudellaro, E.~Piedipalumbo, and S.~Capozziello.
\newblock {Oscillating dark energy: A Possible solution to the problem of eternal acceleration}.
\newblock {\em Phys. Rev. D}, 68:123501, 2003.

\bibitem{Linder:2005dw}
Eric~V. Linder.
\newblock {On oscillating dark energy}.
\newblock {\em Astropart. Phys.}, 25:167--171, 2006.

\bibitem{Feng:2004ff}
Bo~Feng, Mingzhe Li, Yun-Song Piao, and Xinmin Zhang.
\newblock {Oscillating quintom and the recurrent universe}.
\newblock {\em Phys. Lett. B}, 634:101--105, 2006.

\bibitem{Nojiri:2006ww}
Shin'ichi Nojiri and Sergei~D. Odintsov.
\newblock {The Oscillating dark energy: Future singularity and coincidence problem}.
\newblock {\em Phys. Lett. B}, 637:139--148, 2006.

\bibitem{Kurek:2007bu}
Aleksandra Kurek, Orest Hrycyna, and Marek Szydlowski.
\newblock {Constraints on oscillating dark energy models}.
\newblock {\em Phys. Lett. B}, 659:14--25, 2008.

\bibitem{Jain:2007fa}
Deepak Jain, Abha Dev, and Jailson~Souza Alcaniz.
\newblock {Cosmological bounds on oscillating dark energy models}.
\newblock {\em Phys. Lett. B}, 656:15--18, 2007.

\bibitem{Saez-Gomez:2008mkj}
Diego Saez-Gomez.
\newblock {Oscillating Universe from inhomogeneous EoS and coupled dark energy}.
\newblock {\em Grav. Cosmol.}, 15:134--140, 2009.

\bibitem{Kurek:2008qt}
Aleksandra Kurek, Orest Hrycyna, and Marek Szydlowski.
\newblock {From model dynamics to oscillating dark energy parametrisation}.
\newblock {\em Phys. Lett. B}, 690:337--345, 2010.

\bibitem{Pace:2011kb}
F.~Pace, C.~Fedeli, L.~Moscardini, and M.~Bartelmann.
\newblock {Structure formation in cosmologies with oscillating dark energy}.
\newblock {\em Mon. Not. Roy. Astron. Soc.}, 422:1186--1202, 2012.

\bibitem{Pan:2017zoh}
Supriya Pan, Emmanuel~N. Saridakis, and Weiqiang Yang.
\newblock {Observational Constraints on Oscillating Dark-Energy Parametrizations}.
\newblock {\em Phys. Rev. D}, 98(6):063510, 2018.

\bibitem{Panotopoulos:2018sso}
Grigoris Panotopoulos and \'Angel Rinc\'on.
\newblock {Growth index and statefinder diagnostic of Oscillating Dark Energy}.
\newblock {\em Phys. Rev. D}, 97(10):103509, 2018.

\bibitem{Rezaei:2019roe}
Mehdi Rezaei.
\newblock {Observational constraints on the oscillating dark energy cosmologies}.
\newblock {\em Mon. Not. Roy. Astron. Soc.}, 485:550, 2019.

\bibitem{Yao:2022jrw}
Tian-Ying Yao, Rui-Yun Guo, and Xin-Yue Zhao.
\newblock {Constraining neutrino mass in dynamical dark energy cosmologies with the logarithm parametrization and the oscillating parametrization}.
\newblock 11 2022.

\bibitem{Rezaei:2024vtg}
Mehdi Rezaei.
\newblock {Oscillating Dark Energy in Light of the Latest Observations and Its Impact on the Hubble Tension}.
\newblock {\em Astrophys. J.}, 967(1):2, 2024.

\bibitem{Tian:2019enx}
S.~X. Ti\'an.
\newblock {Cosmological consequences of a scalar field with oscillating equation of state: A possible solution to the fine-tuning and coincidence problems}.
\newblock {\em Phys. Rev. D}, 101(6):063531, 2020.

\bibitem{Adil:2023exv}
Shahnawaz~A. Adil, \"Ozg\"ur Akarsu, Eleonora Di~Valentino, Rafael~C. Nunes, Emre \"Oz\"ulker, Anjan~A. Sen, and Enrico Specogna.
\newblock {Omnipotent dark energy: A phenomenological answer to the Hubble tension}.
\newblock {\em Phys. Rev. D}, 109(2):023527, 2024.

\bibitem{Mbonye:2022cnf}
Manasse~R. Mbonye.
\newblock {Is cosmic dynamics self-regulating?}
\newblock {\em Int. J. Mod. Phys. D}, 32(12):2350076, 2023.

\bibitem{Mbonye:2024mss}
Manasse~R. Mbonye.
\newblock {The Big Bang: Origins and initial conditions from Self-Regulating Cosmology (SRC) model}.
\newblock 4 2024.

\bibitem{Zhao:2005vj}
Gong-Bo Zhao, Jun-Qing Xia, Mingzhe Li, Bo~Feng, and Xinmin Zhang.
\newblock {Perturbations of the quintom models of dark energy and the effects on observations}.
\newblock {\em Phys. Rev. D}, 72:123515, 2005.

\bibitem{Heisenberg:2020ywd}
Lavinia Heisenberg, Matthias Bartelmann, Robert Brandenberger, and Alexandre Refregier.
\newblock {Model independent analysis of supernova data, dark energy, trans-Planckian censorship and the swampland}.
\newblock {\em Phys. Lett. B}, 812:135990, 2021.

\bibitem{Haude:2019qms}
Sophia Haude, Shabnam Salehi, Sof\'\i{}a Vidal, Matteo Maturi, and Matthias Bartelmann.
\newblock {Model-Independent Determination of the Cosmic Growth Factor}.
\newblock 12 2019.

\bibitem{Kessler:2025kju}
Daniel~A. Kessler, Luis~A. Escamilla, Supriya Pan, and Eleonora Di~Valentino.
\newblock {One-parameter dynamical dark energy: Hints for oscillations}.
\newblock 4 2025.

\bibitem{Mukhanov:1990me}
Viatcheslav~F. Mukhanov, H.~A. Feldman, and Robert~H. Brandenberger.
\newblock {Theory of cosmological perturbations. Part 1. Classical perturbations. Part 2. Quantum theory of perturbations. Part 3. Extensions}.
\newblock {\em Phys. Rept.}, 215:203--333, 1992.

\bibitem{Mukhanov:1996ak}
Viatcheslav~F. Mukhanov, L.~Raul~W. Abramo, and Robert~H. Brandenberger.
\newblock {On the Back reaction problem for gravitational perturbations}.
\newblock {\em Phys. Rev. Lett.}, 78:1624--1627, 1997.

\bibitem{Brandenberger:2002sk}
Robert~H. Brandenberger.
\newblock {Back reaction of cosmological perturbations and the cosmological constant problem}.
\newblock In {\em {18th IAP Colloquium on the Nature of Dark Energy: Observational and Theoretical Results on the Accelerating Universe}}, 10 2002.

\bibitem{Payeur:2024kyy}
Guillaume Payeur, Evan McDonough, and Robert Brandenberger.
\newblock {Swampland conjectures constraints on dark energy from a highly curved field space}.
\newblock {\em Phys. Rev. D}, 110(10):106011, 2024.

\bibitem{Abramo:1997uy}
L.~Raul~W. Abramo.
\newblock {The Back reaction of gravitational perturbations and applications in cosmology}.
\newblock Other thesis, 9 1997.

\bibitem{Weinberg:2013agg}
David~H. Weinberg, Michael~J. Mortonson, Daniel~J. Eisenstein, Christopher Hirata, Adam~G. Riess, and Eduardo Rozo.
\newblock {Observational Probes of Cosmic Acceleration}.
\newblock {\em Phys. Rept.}, 530:87--255, 2013.

\bibitem{Chen:2024tfp}
Shi-Fan Chen et~al.
\newblock {Baryon Acoustic Oscillation Theory and Modelling Systematics for the DESI 2024 results}.
\newblock 2 2024.

\bibitem{DESI:2024uvr}
A.~G. Adame et~al.
\newblock {DESI 2024 III: Baryon Acoustic Oscillations from Galaxies and Quasars}.
\newblock 4 2024.

\bibitem{Clifton:2024mdy}
Timothy Clifton and Neil Hyatt.
\newblock {A radical solution to the Hubble tension problem}.
\newblock {\em JCAP}, 08:052, 2024.

\bibitem{Ormondroyd:2025iaf}
A.~N. Ormondroyd, W.~J. Handley, M.~P. Hobson, and A.~N. Lasenby.
\newblock {Comparison of dynamical dark energy with \ensuremath{\Lambda}CDM in light of DESI DR2}.
\newblock 3 2025.

\bibitem{Tiwari:2024gzo}
Yashi Tiwari, Ujjwal Upadhyay, and Rajeev~Kumar Jain.
\newblock {Exploring cosmological imprints of phantom crossing with dynamical dark energy in Horndeski gravity}.
\newblock 12 2024.

\bibitem{Colgain:2025nzf}
Eoin~\'O. Colg\'ain, Saeed Pourojaghi, M.~M. Sheikh-Jabbari, and Lu~Yin.
\newblock {How much has DESI dark energy evolved since DR1?}
\newblock 4 2025.

\bibitem{Poulin:2025nfb}
Vivian Poulin, Tristan~L. Smith, Rodrigo Calder\'on, and Th\'eo Simon.
\newblock {Impact of ACT DR6 and DESI DR2 for Early Dark Energy and the Hubble tension}.
\newblock 5 2025.

\bibitem{Pang:2025lvh}
Ye-Huang Pang, Xue Zhang, and Qing-Guo Huang.
\newblock {The Impact of the Hubble Tension on the Evidence for Dynamical Dark Energy}.
\newblock 3 2025.

\bibitem{DESI:2024lzq}
A.~G. Adame et~al.
\newblock {DESI 2024 IV: Baryon Acoustic Oscillations from the Lyman Alpha Forest}.
\newblock 4 2024.

\end{thebibliography}

\end{document}